\documentclass[showpacs, preprintnumbers, prd, nofootinbib, floats, amssymb,
floatfix]{revtex4}
\usepackage{amsmath}
\usepackage{amsxtra}
\usepackage{amssymb}
\usepackage[mathscr]{euscript}
\usepackage{bbold}
\usepackage{bm}
\usepackage{tensor}
\usepackage{blkarray}
\usepackage{graphicx}
\usepackage{xcolor}
\graphicspath{{Imagenes/}}
\usepackage{hyperref}
\usepackage{cleveref}
\begin{document}
	
	\title{
		The Hamilton--Jacobi and Symplectic Analysis
		for Extended Ho\v{r}ava Gravity
	}
	
	\author{Diana Vanessa Castro Luna}
	\email{dcastro@ifuap.buap.mx}
	
	\author{Alberto Escalante}
	\email{aescalan@ifuap.buap.mx}
	
	\affiliation{
		Instituto de F\'isica,
		Benem\'erita Universidad Aut\'onoma de Puebla.\\
		Apartado Postal J-48, 72570 Puebla, Pue., M\'exico
	}
	
	\begin{abstract}
		The Hamilton-Jacobi [HJ] analysis and the Faddeev-Jackiw [FJ] quantization for the linearized, non-projectable $\lambda R$ model, extended by the lowest-order term of Blas, Pujolàs, and Sibiryakov and the higher-order term involving the square of the Cotton tensor, were carried out. The cases $\lambda \neq \frac{1}{3}$ and $\lambda = \frac{1}{3}$ are analyzed. Within the Hamilton-Jacobi framework, we obtain the complete set of Hamiltonians, generalized brackets, fundamental differential, characteristic equations, and gauge transformations of the theory. In the Faddeev-Jackiw framework, a symplectic tensor is constructed from which the fundamental brackets are identified. The HJ and FJ brackets coincide, and from the symplectic tensor, the measure for the quantum path integral is calculated. The differences between the cases $\lambda \neq \frac{1}{3}$ and $\lambda = \frac{1}{3}$ at the quantum path integral are discussed.

	\end{abstract}
	
	\date{\today}
	
	
	\preprint{}
	
	\maketitle
	

	\section{Introduction}
The quest for a consistent quantum theory of gravity has led to various research efforts to unify General Relativity [GR] with quantum mechanics \cite{rovelli1, thiemann1,Kiefer}. From a perturbative perspective, it is well-known that the Einstein-Hilbert [EH] theory is non-renormalizable, but the inclusion of higher-order derivative terms can improve its behavior in the ultraviolet regime. In this respect, there are two ways of adding higher-order derivative terms: spatial or temporal high-order terms. In fact, theories with higher-order time derivatives typically introduce ``ghost" degrees of freedom associated with ordinary curvature squared
theories, leading to instabilities. Consequently, an alternative approach preserves second-order time derivatives while incorporating higher-order spatial derivatives.
Pursuing this idea, Hořava proposed a theory of gravity characterized by an anisotropic treatment of space and time \cite{horava2, horava1, visser}. In four spacetime dimensions, the inclusion of spatial derivatives up to the sixth order corresponds to an anisotropic scaling with a critical exponent of \(z=3\), under the following transformations:	$$
	t\rightarrow b^{z}t,
	\qquad
	x^{i}\rightarrow b x^{i}.
	$$
This construction improves the theory's behavior in the ultraviolet regime and suggests its renormalizability based on a power-counting analysis \cite{bellorin7, Barvinsky1}; furthermore, general relativity is recovered in the low-energy limit through an appropriate choice of coupling constants. The anisotropic treatment of spacetime explicitly breaks Lorentz invariance at high energies and reduces the full group of spacetime diffeomorphisms to the group of foliation-preserving diffeomorphisms [FDiff]. Consequently, Hořava gravity is naturally formulated in terms of the lapse function, the shift vector, and the spatial metric associated with a preferred foliation of spacetime—that is, in ADM variables.\\
Hořava gravity can be formulated in either its projectable or non-projectable version. In the projectable case, the lapse function is restricted to depending solely on time, \(N=N(t)\), whereas in the non-projectable theory, it can depend on both space and time, \(N=N(t,\mathbf{x})\). This distinction has important consequences for  theory's constraint structure and dynamics. In particular, the reduced symmetry compared to general relativity may allow an additional scalar gravitational degree of freedom to propagate. The stability and physical interpretation of this scalar mode depend on the specific version of the theory and the terms included in its potential \cite{bellorin0, bellorin1, bellorin4, bellorin5, bellorin6}.\\
In the non-projectable formulation, the \(\lambda R\) model (the simpler terms of Hořava theory) is obtained by truncating the gravitational potential to its lowest-order contribution in spatial curvature. The parameter \(\lambda\) measures the deviation of the kinetic sector from general relativity. However, the original Hořava theory exhibits pathologies in the scalar mode. In this regard, a consistent extension proposed by Blas, Pujolàs, and Sibiryakov [BPS] incorporates terms built from spatial derivatives of the lapse function. These terms improve the scalar sector and avoid the pathologies of the non-projectable minimal theory \cite{bps}. They modify the constraint structure and play an important role in the scalar sector of the non-projectable theory.  The model can be further extended by including the square of
the Cotton tensor, which is introduced into  Hořava's original potential under the condition of “detailed balance”  \cite{horava1}.  From a renormalization and effective field theory standpoint, it represents a "soft breaking" of anisotropic Weyl symmetry where the UV $z=3$ curvature sector is conformally invariant while conformal symmetry is broken only by lower-derivative IR relevant operators ( we will find to $R$ and $\alpha a_i a^i$). In the ultraviolet, it yields high-order momentum dependence in the propagator, and the absence of higher time derivatives preserves the basic Hořava strategy of improving ultraviolet behavior without introducing the standard higher-time-derivative ghost mechanism. The so-called critical value \(\lambda=1/3\) is particularly relevant because the generalized DeWitt metric becomes degenerate in three spatial dimensions. This gives rise to an additional primary constraint and a constraint structure distinct from the generic case.\\
With the previous points established, we analyze the linearized extended non-projectable $\lambda R$ model in two ways, applying the Hamilton-Jacobi and Faddeev-Jackiw formulations. In fact, the model is extended to include the lowest-order BPS contribution and the Cotton tensor squared term, considering the cases $\lambda \neq 1/3$ and $\lambda = 1/3$ separately. A key aspect of our construction is that we work directly with the metric perturbation, taking the field $h_{\mu\nu}$ as the dynamical variable and performing a $(3+1)$ decomposition, rather than perturbing the ADM variables from the outset; we have shown that both procedures yield equivalent results at the perturbative level in the \(\lambda R\) model  \cite{diana1}. Previous canonical studies of this extended model have used the Dirac-Bergmann algorithm \cite{fer2} . In this formalism, the complete set of constraints is classified into first- and second-class constraints based on their Poisson bracket algebra \cite{Dirac,teitelboim}, but a complete study of the constraints was not presented. Although this procedure provides a systematic description of singular systems, the implementation of consistency conditions and the classification of the resulting constraints can be technically complex. Consequently, the Hamilton-Jacobi formulation offers a complementary perspective on the canonical structure of the extended model.\\
Within the Hamilton-Jacobi framework for singular systems \cite{guller1,guller2,guller3,guller4}, constraints are introduced as Hamiltonians, and the consistency of the resulting system is examined via Frobenius integrability conditions. These conditions can generate additional Hamiltonians until the complete constraint structure is obtained. Subsequently, the Hamiltonians are classified as involutive or non-involutive based on their mutual Poisson brackets. When non-involutive Hamiltonians exist, they are incorporated into the dynamics through generalized Hamilton-Jacobi brackets, thereby reducing the system to an involutive one.\\
Once the system is integrable, the involutive Hamiltonians define the fundamental differential of the theory, from which the characteristic equations governing the evolution of the canonical variables are derived. Furthermore, by setting $dt=0$ in the characteristic equations, it is possible to obtain the gauge transformations; this allows for a direct examination of the theory's invariance under such transformations.\\
In contrast, the Faddeev-Jackiw method focuses on constructing a symplectic tensor. This symplectic approach elegantly gathers all the essential details of the theory within its framework. This is achieved by crafting an invertible symplectic tensor from variables that serve as the theory’s degrees of freedom. Since the theory is singular, constraints naturally emerge. The FJ framework addresses all these constraints on equal footing, removing the need to sort them into categories like primary, secondary, first class, or second class, as required in Dirac’s method \cite{fad1, fad2, fad3, fad4}. Once the symplectic tensor is obtained, its components correspond to the FJ generalized brackets. Ultimately, the brackets defined by Dirac and those by FJ are the same. In this respect, we report a complete analysis of extended \(\lambda R\) model constraints to make progress in the study of the full  Hořava's  theory.  \\
The paper is structured as follows: Section II introduces the extended, linearized $\lambda R$ model, which forms the basis of our analysis. Sections III and IV deal with the Hamilton-Jacobi formulation of Hořava's action, considering the cases when $\lambda \neq 1/3$ and when $\lambda = 1/3$, respectively. We determine all the Hamiltonians of the theory and derive the associated characteristic equations. We then find the gauge transformations generated by the involutive Hamiltonians and investigate the action  invariance under these transformations. In Sections VI and VII, the symplectic analysis is carried out, a symplectic tensor is constructed, and the results from the Hamilton-Jacobi approach are reproduced. Section VIII contains the conclusions.

	\section{Linearized $\lambda R$ gravity plus the BPS and a Cotton-square terms}

	The action for $\lambda R$ model in the ADM formulation is described by the following action
	
	\begin{equation}
	\label{accion1}
		S = \int dt d^{3}x \sqrt{g} N \left( G^{ijkl} K_{ij} K_{kl} + R\right),
	\end{equation}
	where $	K_{ij} = \frac{1}{2N} \left( \dot{g}_{ij} - 2 \nabla_{(i}N_{j)} \right)$ is the extrinsic curvature, $N$ is the lapse function, $g_{ij}$ is the spatial metric, $R$ is the spatial Ricci scalar and $G^{ijkl}$ is a generalization of the DeWitt metric defined by
	
	\begin{equation}
		G^{ijkl} = \frac{1}{2} \left( g^{ik} g^{jl} + g^{il} g^{jk} \right) - \lambda g^{ij} g^{kl}.
	\end{equation}
	The parameter $\lambda$ characterizes the deviation of the $\lambda R$ model from General Relativity [GR]. In the limit when $\lambda \rightarrow 1$, GR is recovered in ADM formulation. Moreover, the action (\ref{accion1}) is invariant under foliation-preserving diffeomorphisms [FDiff]. In this study, rather than beginning with the action (\ref{accion1}), we start with the well-known Fierz-Pauli action and introduce the parameter $\lambda$ to build the perturbative $\lambda R$ model, this is

	\begin{equation*}\label{actionlingravity}
		S = \int d^{4}x \left(  \frac{1}{4} \partial_{\lambda} h_{\mu \nu} \partial^{\lambda} h^{\mu \nu } -   \frac{1}{4} \partial_{\lambda} h_{\mu}^ {\mu} \partial^{\lambda} h^{\nu}_ {\nu } +  \frac{1}{2} \partial_{\lambda} h^{\lambda}_{\mu} \partial^{\mu} h_{\nu}^{\nu} - \frac{1}{2} \partial_{\lambda} h^{\lambda}_{\mu} \partial_{\nu} h^{\mu \nu}   \right),
	\end{equation*}
	where we have considered  a perturbative approach around a Minkowski background. More specifically, by performing the  $3+1$ decomposition 
	
	\begin{equation}\label{FPfinal}
		\begin{split}
			\mathscr{L}_{FP} & = \frac{1}{4} \dot{h}_{ij} \dot{h}^{ij} - \dot{h}^{ij} \partial_{i} h_{0j}  - \dot{h}^{j}_{j} \partial_{i} h^{0i} - \frac{1}{4} (\dot{h}^{i}_{i} )^{2} - \frac{1}{2} \partial_{i} h_{0j} \partial^{i} h^{0j} \\
			&+ \frac{1}{2} \partial^{i} h^{j 0} \partial_{j} h_{i0} + \frac{1}{2} \partial_{i} h_{00} \partial^{j} h^{ij} - \frac{1}{2} \partial_{i} h_{k}^{k} \partial_{j} h^{ij} - \frac{1}{2} \partial_{i} h_{00} \partial^{i}  h_{k}^{k} \\
			&+ \frac{1}{4} \partial_{i} h_{j}^{j} \partial^{i} h_{k}^{k} + \frac{1}{2} \partial^{i} h^{jk} \partial_{j} h_{ik} - \frac{1}{4} \partial_{i} h_{jk} \partial^{i} h^{jk},
		\end{split}
	\end{equation}
	where the perturbation is given by $g_{\mu \nu} = \eta_{\mu \nu} + h_{\mu \nu}$ with $\eta_{\mu \nu} = diag\left(-1, +1, +1, +1\right) $. We further introduce the following change of variables for rewriting   \eqref{FPfinal} in a $\lambda R$-like model
	
	\begin{equation}
		K_{ij} = \frac{1}{2} \left( \dot{h}_{ij} - \partial_{i} h_{0j} - \partial_{j}h_{0i} \right), \label{kij} \\
	\end{equation}
	although this quantity is not dynamical, it provides a useful tool for rewriting the Fierz--Pauli Lagrangian in a form suitable for introducing the $\lambda R$ model
	
	\begin{equation}
		\mathscr{L}_{FP} = G^{ijkl} K_{ij} K_{kl} - \frac{1}{2} h^{00}R_{ij}^{ij} - \frac{1}{2} h_{ij} \left( R_{ikj}^{k} - \frac{1}{2} \delta_{ij} R_{lm}^{lm} \right),
	\end{equation}
	where
	
	\begin{align}
		R_{ij} & = \frac{1}{2} \left( \partial_{k} \partial_{i} h_{j}^{k}  - \partial^{k} \partial_{k} h_{ij} - \partial_{j} \partial_{i} h_{k}^{k} + \partial_{j} \partial^{k} h_{ik} \right),\\
		R_{ij}^{ij} & = \partial^{i} \partial^{j} h_{ij} - \nabla^{2} h,
	\end{align}
	and
	
	\begin{equation}
		G^{ijkl} = \frac{1}{2} \left( \delta^{ik} \delta^{jl} + \delta^{il} \delta^{jk} \right) - \lambda \delta^{ij} \delta^{kl}. \label{dewittmetric}
	\end{equation}
	The $\lambda R$ model serves as a starting point for the construction considered in this work, as it corresponds to the lowest-order truncation of the potential sector of non-projectable Hořava gravity. This truncation provides a basis for extending the model by incorporating the BPS term \cite{bps}. Such an extension is justified by the structure of non-projectable Hořava gravity, which is based on the group of foliation-preserving diffeomorphisms (FDiff). In the full theory, the reduced symmetry allows for the existence of an additional scalar degree of freedom, alongside the two degrees of freedom present in General Relativity.\\
	The extended $\lambda R$ model is obtained by including the BPS term with $z=1$ \cite{bps}. This term is compatible with foliation-preserving diffeomorphisms and is given by the expression $a_i=\partial_i\ln N$, which is covariant under FDiff. The resulting model constitutes the lowest-order effective action of the full Hořava theory, up to second order in spatial derivatives.\\
	As a further extension, the squared Cotton term—derived from the detailed balance condition \cite{horava1}—is incorporated into the effective potential. This allows us to consider a Hořava theory with soft breaking of conformal symmetry \cite{bellorin2,bellorin3, park}. In this construction, the extended $\lambda R$ sector is not conformally invariant, whereas the contribution from the squared Cotton term is. The Cotton tensor is defined as
	
	\begin{equation}
		C^{ij}= \epsilon^{ikl} \nabla_{k} \left( R^{j}_{l} - \frac{1}{4} R \delta^{j}_{l} \right).
	\end{equation}
	By combining the  $\lambda R$ model with the contribution of the square of the Cotton tensor and considering perturbations around a Minkowski background—that is, by linearizing the BPS terms and the square of the Cotton tensor—we obtain the following action
	
	\begin{equation}\label{accionS}
		\begin{split}
			S & = \int \left( G^{ijkl} K_{ij} K_{kl} - \frac{1}{2} h^{00}R_{ij}^{ij} - \frac{1}{2} h_{ij} \left( R_{ij} - \frac{1}{2} \delta_{ij} R \right) + \alpha \partial_{i} h_{00} \partial^{i} h^{00} - \omega \partial_{i} R^{j}_{k} \partial^{i} R^{k}_{j} \right. \\
			& \left.  + \omega \partial^{i} R^{j}_{i} \partial_{k} R^{k}_{j} + \frac{3 \omega}{8}  \partial_{i} R \partial^{i} R +  \frac{\omega}{2} \partial_{i} R^{i}_{j} \partial^{j} R \right) d^{4}x.
		\end{split}
	\end{equation}
	This leads to the following Lagrangian density
	
	\begin{equation}\label{Lag1}
		\begin{split}
			\mathscr{L} & =  G^{ijkl} K_{ij} K_{kl} - \frac{1}{2} h^{00}R_{ij}^{ij} - \frac{1}{2} h_{ij} \left( R_{ij} - \frac{1}{2} \delta_{ij} R \right) + \alpha \partial_{i} h_{00} \partial^{i} h^{00} - \omega \partial_{i} R^{j}_{k} \partial^{i} R^{k}_{j} \\
			& + \omega \partial^{i} R^{j}_{i} \partial_{k} R^{k}_{j} + \frac{3 \omega}{8}  \partial_{i} R \partial^{i} R +  \frac{\omega}{2} \partial_{i} R^{i}_{j} \partial^{j} R,
		\end{split}
	\end{equation}
	where $\alpha$ and $\omega$ are coupling constants of the BPS term  and the square of the Cotton tensor, respectively.

	\section{HJ analysis for $\lambda \neq \frac{1}{3}$}
	
	Now we  perform the Hamilton--Jacobi [HJ] formalism for the action (\ref{Lag1}). First, we  introduce the canonical momenta, given by
	
	\begin{align}
		\pi^{00} & = \frac{ \partial \mathscr{L}}{\partial h_{00}} = 0,\\
		\pi^{0i} &  = \frac{ \partial \mathscr{L}}{\partial h_{0i}} = 0,\\
		\pi^{ij} & = \frac{ \partial \mathscr{L}}{\partial h_{ij}} = G^{ijkl}K_{kl} \label{pij}.
	\end{align}
	The Hamiltonian obtained from the  Legendre transformation is
	
	\begin{equation}
		\begin{split}
			\mathscr{H}_{0} & = \int \left( G_{ijkl}  \pi^{kl} \pi^{ij} - 2 h_{j0} \partial_{i} \pi^{ij}  + \frac{1}{2} h_{00} R +  \frac{1}{2}  h^{ij}  \left( R_{ij}  - \frac{1}{2}  \delta_{ij} R \right) -  \alpha \partial_{i} h_{00} \partial^{i} h^{00} \right.  \\
			& \left. + \omega \partial_{i} R^{j}_{k} \partial^{i} R^{k}_{j} -  \omega \partial^{i} R^{j}_{i} \partial_{k} R^{k}_{j} - \frac{3 \omega}{8}  \partial_{i} R \partial^{i} R - \frac{\omega}{2} \partial_{i} R^{i}_{j} \partial^{j} R\right) d^{3}x.
		\end{split}
	\end{equation}
	For this expression, we assume that $\lambda \neq \frac{1}{3}$, since the inverse DeWitt metric
	$G_{ijkl} = \frac{1}{2} \left( \delta_{ik} \delta_{jl} + \delta_{il} \delta_{jk} \right) - \frac{\lambda}{1 - 3 \lambda} \delta_{ij} \delta_{kl}$
	must be well defined. This condition leads us to consider two distinct cases in order to obtain a well-defined Hamiltonian.\\
	From the definition of the momenta, we identify the fundamental Poisson brackets, given by 
	\begin{equation}
		\left\lbrace h_{\mu \nu}(x), \pi^{\alpha \beta}(y) \right\rbrace = \frac{1}{2} \left( \delta^{\alpha}_{\mu} \delta^{\beta}_{\nu} + \delta^{\alpha}_{\nu} \delta^{\beta}_{\mu} \right) \delta^{3}(x-y),
	\end{equation}
	thus, we can write  the following canonical brackets
	
	\begin{align}
		\left\lbrace h_{00}(x), \pi^{00}(y) \right\rbrace & = \delta^{3}(x-y),\\
		\left\lbrace h_{0i}(x), \pi^{0j}(y) \right\rbrace & =\frac{1}{2}  \delta^{j}_{i} \delta^{3}(x-y),\\
		\left\lbrace h_{ij}(x), \pi^{kl}(y) \right\rbrace & = \frac{1}{2} \left( \delta^{k}_{i} \delta^{l}_{j} + \delta^{k}_{j} \delta^{l}_{i} \right) \delta^{3}(x-y).
	\end{align}
	We have mentioned that, in the Hamilton-Jacobi formalism, the constraints are referred to as Hamiltonians. It is also worth noting that all canonical variables are treated on the same footing. For this model, the initial set of Hamiltonians is given by
	
	\begin{align*}
		H_{0} & = G_{ijkl}  \pi^{kl} \pi^{ij} - 2 h_{j0} \partial_{i} \pi^{ij}  + \frac{1}{2} h_{00} R +  \frac{1}{2}  h_{ij} R_{ij}  - \frac{1}{4} hR-  \alpha \partial_{i} h_{00} \partial^{i} h^{00} \\
		&+ \omega \partial_{i} R_{jk} \partial^{i} R_{kj} -  \omega \partial^{i} R_{ji} \partial_{k} R_{kj} - \frac{3 \omega}{8}  \partial_{i} R \partial^{i} R - \frac{\omega}{2} \partial_{i} R_{ij} \partial^{j} R,\\
		H_{1} & = \pi^{00},\\
		H^{i}_{2} & = \pi^{0i}.
	\end{align*}
	To determine the integrability of the system, the Hamiltonians must be classified as involutive or non-involutive. To this end, it is necessary to calculate their mutual Poisson brackets. The Hamiltonians are involutive if their Poisson brackets vanish or close on the set of Hamiltonians; otherwise, they are non-involutive. The primary Hamiltonians satisfy $\left\lbrace H_{1}, H^{i}_{2} \right\rbrace = 0$. Therefore, the initial fundamental differential can be written as
	
	\begin{equation}
		df = \int \left[ \left\lbrace f, H'_{0} \right\rbrace dt +\left\lbrace f, H_{1} \right\rbrace d\omega_{1} + \left\lbrace f, H^{i}_{2} \right\rbrace d\omega_{2i}  \right] d^{3}x.
	\end{equation}
	The Hamiltonian $H'_{0}$ is defined as $H'_{0} \equiv H_{0} + \partial_{0} S$, where $S$ is the action  and the $\omega_{a}$ are evolution parameters. Following the Hamilton-Jacobi formalism, we require the Hamiltonians to satisfy the Frobenius integrability conditions, $dH_{a}=0$ —where $H_{a}=(H_{1},H_{2}^{i})$—, to ensure that the system is in involution. Applying these conditions to the initial set of Hamiltonians yields the following additional Hamiltonians:
	
	\begin{align}
		H_{3} & =  \frac{1}{2} R  + 2 \alpha \nabla^{2} h_{00},\\
		H^{i}_{4}  & = \partial_{j} \pi^{ij}.
	\end{align}
	The brackets among the resulting Hamiltonians are
	
	\begin{align}
		\left\lbrace H^{i}_{2}(x), H^{j}_{4}(y) \right\rbrace & = 0, \\
		\left\lbrace H_{3}(x), H_{1}(y) \right\rbrace &  = 2 \alpha \nabla_{x}^{2} \delta^{3}(x-y).
	\end{align}
	Therefore, $H^{i}_{2}$ and $H^{i}_{4}$ form an involutive subset, whereas $H_{1}$ and $H_{3}$ constitute a non-involutive pair. Given that $H_{1}$ and $H_{3}$ are non-involutive, they cannot be retained in the involutive set under the original bracket; thus, to remove them, the HJ bracket is redefined as
	
	\begin{equation}\label{newbra}
		\left\lbrace f(x), g(y) \right\rbrace^{*} = 	\left\lbrace f(x), g(y) \right\rbrace - \int 	\left\lbrace f(x), H^{ni}_{a}(u) \right\rbrace C^{-1}_{ab} \left\lbrace H^{ni}_{b}(v), g(y) \right\rbrace du dv,
	\end{equation}
	where $H^{ni}$ are non-involuting Hamiltonians, $f$ and $g$ are phase space functions, and the matrix $C_{ab}$ is that whose entries are the Poisson brackets between non-involuting Hamiltonians, and is given by
	
	\begin{equation}
		C_{ab} = 2\alpha
		\left(
		\begin{array}{cc}
			0 & -1 \\ 
			1 & 0  \\
		\end{array}
		\right) \nabla_{u}^{2} \delta^{3}(u-v),
	\end{equation}
	and its inverse is
	
	\begin{equation}
		(C_{ab})^{-1} = \frac{1}{2\alpha \nabla_{u}^{2}}
		\left(
		\begin{array}{cc}
			0 & 1 \\ 
			-1 & 0  \\
		\end{array}
		\right) \delta^{3}(u-v).
	\end{equation}
	Using the new HJ bracket \eqref{newbra}, we obtain the following fundamental brackets:
	
	\begin{align}
		\left\lbrace h_{00}(x), \pi^{00}(y) \right\rbrace^{*} & = 0,\\
		\left\lbrace h_{00}(x), \pi^{ij}(y) \right\rbrace^{*} & =\frac{1}{4 \alpha \nabla^{2} } P^{ij} \delta^{3} (x-y),\\
		\left\lbrace h_{0i}(x), \pi^{0j}(y) \right\rbrace^{*} & = \frac{1}{2}  \delta^{i}_{j} \delta^{3}(x-y),\\
		\left\lbrace h_{ij}(x), \pi^{kl}(y) \right\rbrace^{*} & = \frac{1}{2} \left( \delta^{k}_{i} \delta^{l}_{j} + \delta^{k}_{j} \delta^{l}_{i} \right) \delta^{3}(x-y),
	\end{align}
	where $P^{ij}$ is defined by $P^{ij} = \partial^{i} \partial^{j} - \delta^{ij} \nabla^{2} $. After eliminating the non-involutive Hamiltonians, we are left with the following set:
	
	\begin{align*}
		H_{0} & = G_{ijkl}  \pi^{kl} \pi^{ij} - 2 h_{j0} \partial_{i} \pi^{ij}  + \frac{1}{2} h_{00} R +  \frac{1}{2}  h_{ij} R_{ij}  - \frac{1}{4} hR-  \alpha \partial_{i} h_{00} \partial^{i} h^{00} \\
		&+ \omega \partial_{i} R_{jk} \partial^{i} R_{kj} -  \omega \partial^{i} R_{ji} \partial_{k} R_{kj} - \frac{3 \omega}{8}  \partial_{i} R \partial^{i} R - \frac{\omega}{2} \partial_{i} R_{ij} \partial^{j} R, \\
		H^{i}_{2} & = \pi^{0i},\\
		H^{i}_{4}  & = \partial_{j} \pi^{ij},
	\end{align*}
	and the new fundamental differential now reads 
	
	\begin{equation}\label{fd1}
		df  = \int \left[ \left\lbrace f, H_{0} \right\rbrace^{*} dt + \left\lbrace f, H^{i}_{2} \right\rbrace^{*} d\omega_{2i}  + \left\lbrace f, H^{i}_{4} \right\rbrace^{*} d\omega_{4i} \right] d^{3}x.\\
	\end{equation}
	The system is involutive since the integrability conditions are satisfied. Using the new fundamental differential, we obtain
	
	\begin{equation}
		\begin{split}
			dH^{i}_{2} & = H^{i}_{4}  = \partial_{j} \pi^{ij},\\
			dH^{i}_{4} & = 0.
		\end{split}
	\end{equation}
	Now, we can compute the characteristic equations for the canonical variables  using \eqref{fd1} which gives
	
	\begin{align}
		dh_{00} 	& =  \frac{1}{2 \alpha \nabla^{2}}   P^{ij} G_{ijkl}  \pi^{kl} dt =  \frac{1}{2 \alpha}   \frac{\lambda - 1 }{1 - 3 \lambda} \pi dt ,\\
		dh_{0i} & = \frac{1}{2} d\omega_{2i}, \\
		dh_{ij} & = \left( 2 K_{ij}  + \partial_{i} h_{0j}  +  \partial_{j}h_{0i} \right) dt - \frac{1}{2} \left( \delta^{m}_{i} \partial_{j} + \delta^{m}_{j} \partial_{i} \right) d\omega_{4m}, \\
		d\pi^{00} &= 0, \\
		d\pi^{0i} & = \partial_{j} \pi^{ij} dt, \\
		d \pi^{ij} & = \left( - \frac{1}{8 \alpha \nabla^{2}} P^{ij}  R - P^{ij} h_{00} - R^{ij} + \frac{1}{2} \delta^{ij} R - \omega \nabla^{4} R^{ij} - \frac{3}{4} \omega \nabla^{2} \partial^{i}\partial^{j}R  \right. \\
		& \left.+\frac{5}{4} \omega \delta^{ij}\nabla^{4}R \right) dt. \nonumber 
	\end{align}
	These equations describe the system's physical dynamics. We observe that the variables  $h_{0i}$ are nondynamical, as they are associated with the Lagrange multipliers and the involutive Hamiltonians. In contrast to the results reported in \cite{diana1}, the counting of  physical degrees of freedom in the present case is carried out in a different way. The variables $h_{ij}$ and $\pi^{ij}$ provide 12 dynamical variables, while the six involutive Hamiltonians $H^{i}_{2}$ and $H^{i}_{4}$ remove six degrees of freedom. Therefore,
	
	\begin{equation}
		DoF = \frac{1}{2} (12-6) = 3.
	\end{equation}
	All these results  extend  those reported in \cite{fer2}.
	
	\section{HJ analysis for $\lambda = \frac{1}{3}$}
	
	The case $\lambda = \frac{1}{3}$ requires a separate analysis because the generalized DeWitt metric becomes degenerate. In particular, its inverse is not well-defined, and the canonical momentum $\pi^{ij}$ satisfies an additional primary constraint. For $\lambda = \frac{1}{3}$, the canonical momenta are
	
	\begin{align}
		\pi^{00} & = \frac{ \partial \mathscr{L}}{\partial h_{00}} = 0,\\
		\pi^{0i} &  = \frac{ \partial \mathscr{L}}{\partial h_{0i}} = 0,\\
		\pi^{ij} & = \frac{ \partial \mathscr{L}}{\partial h_{ij}} = G^{ijkl}K_{kl} \label{pij}.
	\end{align}
	In this case, the generalized DeWitt metric becomes
	$G^{ijkl} = \frac{1}{2} \left( \delta^{ik} \delta^{jl} + \delta^{il} \delta^{jk} \right) - \frac{1}{3} \delta^{ij} \delta^{kl}$. Furthermore, by considering the trace of $\pi^{ij}$, we obtain
	
	\begin{equation}
		\begin{split}
			\pi^{ij}  & = \frac{ \partial \mathscr{L}}{\partial h_{ij}} =K^{ij} - \frac{1}{3} \delta^{ij} K\\
			\delta_{ij} \pi^{ij}  & =\delta_{ij} K^{ij} - \frac{1}{3} \delta_{ij} \delta^{ij} K\\
			\pi & = 0.
		\end{split}
	\end{equation}  
	The corresponding canonical Hamiltonian is
	
	\begin{equation}
		\begin{split}
			\mathscr{H}_{0} & = \int \left( \pi^{ij} \pi_{ij} - 2 h_{j0} \partial_{i} \pi^{ij}  + \frac{1}{2} h_{00} R_{ij}^{ij} +  \frac{1}{2}  h_{ij}  \left( R^{ij}  - \frac{1}{2}  \delta_{ij} R_{ij}^{ij} \right) -  \alpha \partial_{i} h_{00} \partial^{i} h^{00} \right.  \\
			& \left. + \omega \partial_{i} R^{j}_{k} \partial^{i} R^{k}_{j} -  \omega \partial^{i} R^{j}_{i} \partial_{k} R^{k}_{j} - \frac{3 \omega}{8}  \partial_{i} R \partial^{i} R - \frac{\omega}{2} \partial_{i} R^{i}_{j} \partial^{j} R\right) d^{3}x.
		\end{split}
	\end{equation}
	Thus, the degeneracy of the DeWitt metric gives rise to an extra Hamiltonian; we then have
	
	\begin{align}
		H_{0} & =\pi^{ij} \pi_{ij} - 2 h_{j0} \partial_{i} \pi^{ij}  + \frac{1}{2} h_{00} R +  \frac{1}{2}  h_{ij} R_{ij}  - \frac{1}{4} hR-  \alpha \partial_{i} h_{00} \partial^{i} h^{00} \\
		&+ \omega \partial_{i} R_{jk} \partial^{i} R_{kj} -  \omega \partial^{i} R_{ji} \partial_{k} R_{kj} - \frac{3 \omega}{8}  \partial_{i} R \partial^{i} R - \frac{\omega}{2} \partial_{i} R_{ij} \partial^{j} R, \\
		H_{1} & = \pi^{00},\\
		H^{i}_{2} & = \pi^{0i},\\
		H_{3}& = \pi.
	\end{align}
	The initial Hamiltonians are mutually involutive, and the fundamental differential is therefore given by
	
	\begin{equation}
		df = \int \left[ \left\lbrace f, H_{0} \right\rbrace dt +\left\lbrace f, H_{1} \right\rbrace d\omega_{1} + \left\lbrace f, H^{i}_{2i} \right\rbrace d\omega_{2} +\left\lbrace f, H_{3} \right\rbrace d\omega_{3} \right] d^{3}x.
	\end{equation}
	Imposing the Frobenius integrability conditions on this set yields the following additional Hamiltonians:
	
	\begin{align}
		H_{4} & = \frac{1}{2} R  + 2 \alpha \nabla^{2} h_{00},\\
		H^{i}_{5}  & = \partial_{j} \pi^{ij},\\
		H_{6} & =  \nabla^{2} h_{00} + \frac{1}{2} R + 2 \omega \nabla^{2} \nabla^{2} R.
	\end{align}
	Evaluating the brackets among the complete set of Hamiltonians, we find that $H_1$, $H_3$, $H_4$, and $H_6$ form the non-involutive sector. We therefore introduce the generalized HJ bracket defined in \eqref{newbra}.\\
	The corresponding matrix $C_{ab}$ is
	\begin{equation}
		C_{ab} =
		\left(
		\begin{array}{cccc}
			0 & 0 & - 2\alpha \nabla_{u}^{2} & -\nabla_{u}^{2}  \\ 
			0 & 0 & \nabla_{u}^{2} & \nabla_{u}^{2} + 4 \omega \nabla_{u}^{2} \nabla_{u}^{2} \nabla_{u}^{2}  \\ 
			2\alpha \nabla_{u}^{2} &  - \nabla_{u}^{2}  & 0 & 0\\ 
			\nabla_{u}^{2} & -  \nabla_{u}^{2} - 4 \omega \nabla_{u}^{2} \nabla_{u}^{2} \nabla_{u}^{2}& 0 & 0  \\ 
		\end{array}
		\right) \delta^{3}(u-v),
	\end{equation}
	whose inverse is
	
	\begin{equation}
		C^{-1}_{ab} =
		\left(
		\begin{array}{cccc}
			0 & 0 & -1 - 4 \omega \nabla_{u}^{4}  & 1 \\ 
			0 & 0 & -1  &  2 \alpha \\ 
			1 + 4 \omega \nabla_{u}^{4}  & 1 & 0 & 0\\ 
			-1 & - 2 \alpha & 0 & 0  \\ 
		\end{array}
		\right) \frac{1}{\nabla^{2}_{u} \left( 1 - 2 \alpha - 8 \alpha \omega \nabla^{4}_{u}\right) }\delta^{3}(u-v).
	\end{equation}
	Thus, the non-vanishing fundamental brackets take the form

	\begin{align}
		\left\lbrace h_{00}(x), \pi^{00}(y) \right\rbrace^{*} & = 0,\\
		\left\lbrace h_{00}(x), \pi^{mn}(y) \right\rbrace^{*} & = 0,\\
		\left\lbrace h_{ij}(x), \pi^{00}(y) \right\rbrace^{*} & = 0,\\
		\left\lbrace h_{0i}(x), \pi^{0j}(y) \right\rbrace^{*} & =\frac{1}{2}  \delta^{j}_{i} \delta^{3}(x-y),\\
		\left\lbrace h_{ij}(x), \pi^{mn}(y) \right\rbrace^{*} & =  \frac{1}{2} \left( \delta^{m}_{i} \delta^{n}_{j} + \delta^{m}_{j} \delta^{n}_{i} \right) \delta^{3}(x-y) +  \frac{\delta_{ij} }{2 \nabla^{2}} P^{mn} \delta^{3}(x-y).
	\end{align}
	A significant difference compared to the case $\lambda \neq \frac{1}{3}$ is that we no longer have an explicit dependence on $\alpha$ or $\omega$ within the parentheses. After removing the non-involutive Hamiltonians, the remaining involutive set is
	
	\begin{align*}
		H_{0} & =\pi^{ij} \pi_{ij} - 2 h_{j0} \partial_{i} \pi^{ij}  + \frac{1}{2} h_{00} R +  \frac{1}{2}  h_{ij} R_{ij}  - \frac{1}{4} hR-  \alpha \partial_{i} h_{00} \partial^{i} h^{00} \\
		&+ \omega \partial_{i} R_{jk} \partial^{i} R_{kj} -  \omega \partial^{i} R_{ji} \partial_{k} R_{kj} - \frac{3 \omega}{8}  \partial_{i} R \partial^{i} R - \frac{\omega}{2} \partial_{i} R_{ij} \partial^{j} R, \\
		H^{i}_{2} & = \pi^{0i},\\
		H^{i}_{5}  & = \partial_{j} \pi^{ij}.
	\end{align*}
	The corresponding fundamental differential is
	
	\begin{equation}\label{fd2}
		df  = \int \left[ \left\lbrace f, H_{0} \right\rbrace^{*} dt + \left\lbrace f, H^{i}_{2} \right\rbrace^{*} d\omega_{2}  + \left\lbrace f, H^{i}_{5} \right\rbrace^{*} d\omega_{5} \right] d^{3}x.\\
	\end{equation}
	The integrability conditions are now satisfied
	
	\begin{equation}
		\begin{split}
			dH^{i}_{2} & = H^{i}_{5}  = \partial_{j} \pi^{ij},\\
			dH^{i}_{5} & = 0.
		\end{split}
	\end{equation}
	In this manner, the characteristic equations are consequently obtained as
	
	\begin{align}
		dh_{00} & =  0,\\
		dh_{0i} & = \frac{1}{2} d\omega_{2i}, \\
		dh_{ij} & = \left[ 2 \pi^{ij} - \delta_{ij} \pi+ \partial_{j} h_{0i} + \partial_{i} h_{0j} \right] dt - \frac{1}{2} \left( \delta^{m}_{i} 	\partial_{j}   + \delta^{m}_{j} \partial_{i}  \right) d\omega_{5m}, \\
		d\pi^{00} & = 0,\\
		d\pi^{0i} & = \partial_{j}\pi^{ij} dt, \\
		d\pi^{ij}& =  \left[  \frac{1}{4 \nabla^{2}} P^{ij} R - R^{ij} +  \frac{1}{2} \delta^{ij} R  - \omega \nabla^{4} R^{ij} + \frac{1}{4} \omega \nabla^{2} \partial^{i} \partial^{j} R \right.  \label{traceless} \nonumber  \\ 
		& \left. + \frac{1}{4} \delta^{ij} \omega \nabla^{4} R  \right] dt. 
	\end{align}
	As in the $\lambda\neq\frac{1}{3}$ case, the variables $h_{00}$, $h_{0i}$, $\pi^{00}$, and $\pi^{0i}$ are non-dynamical, as they are associated with the Lagrange multipliers and the involutive Hamiltonians. We can take the trace of \eqref{traceless} and obtain
	
	\begin{equation}
		\dot{\pi}^{i}_{i} = 0.
	\end{equation}
	Therefore, although $h_{ij}$ and $\pi^{ij}$ initially provide 12 phase-space variables, the above result reduces this number to 10. Together with the six involutive Hamiltonians $H_2^i$ and $H_5^i$, the number of physical degrees of freedom is
	
	\begin{equation}
		DoF = \frac{1}{2} (10-6) = 2.
	\end{equation}
	This result differs from the $\lambda\neq\frac{1}{3}$ case, for which three physical degrees of freedom are obtained. The two results are in agreement with the canonical analysis performed using the Dirac formalism in \cite{fer2}.
	
	\section{Gauge transformations}
	
	In the HJ formalism, the characteristic equations encode both the time evolution and the canonical transformations of the system. The latter can be obtained by setting $dt=0$, for both values of $\lambda$, we obtain
	
	\begin{align}
		\delta h_{00} & = 0,\\
		\delta h_{0i} & = \frac{1}{2}  \delta \omega_{2i}, \\
		\delta h_{ij} & =  - \frac{1}{2} \left( \delta^{m}_{i} \partial_{j} + \delta^{m}_{j} \partial_{i}  \right) \delta \omega_{4m}. 
	\end{align}
	Combining the above transformations into a covariant form expression, we can write 
	
	\begin{equation}\label{deltahmunu}
		\delta h_{\mu \nu} = \frac{1}{2} \left( \delta^{0}_{\mu} \delta^{i}_{\nu} + \delta^{0}_{\nu} \delta^{i}_{\mu} \right) \delta \omega_{2i} - \frac{1}{2} \delta^{i}_{\mu} \delta^{j}_{\nu} \left(  \partial_{i} \delta \omega_{4j} + \partial_{j} \delta \omega_{4i} \right).
	\end{equation}
	We need the variation of the action (\ref{accionS}); to this end, we rewrite the action as:
	
	\begin{equation}
		S = S_{\lambda R}+ S_{BPS} + S_{Cotton}.
	\end{equation}
	That is, we decompose the action into the $\lambda R$, BPS, and Cotton-squared sectors. Consequently, its variation is given by the sum of the corresponding contributions,
	
	\begin{equation}
		\begin{split}
			\delta S & = \int d^{4} x \left[ \delta S_{\lambda R} + \delta S_{BPS}  + \delta S_{Cotton}  \right] \\
			& = \int d^{4} x \delta S_{\lambda R} + \int d^{4} x \delta S_{BPS}  + \int d^{4} x \delta S_{Cotton}  \\
			& = \frac{1}{2} \int d^{4}x \left(   - \partial_{\lambda} \partial^{\lambda} h^{\mu \nu} + \eta^{\mu \nu} \partial_{\lambda} \partial^{\lambda} h^{\rho}_{\rho} -  \partial^{\mu} \partial^{\nu} h^{\lambda}_{\lambda} - \eta^{\mu \nu}  \partial^{\lambda} \partial^{\rho} h_{\lambda \rho} \right. \\
			& \left. +  \partial^{\mu} \partial^{\lambda}  h^{\nu}_{\lambda} +\partial^{\nu} \partial^{\lambda} h^{\mu}_{\lambda} \right) \delta h_{\mu \nu}   -2 \alpha \int d^{4}x \left[ \partial_{i}  \partial^{i} h^{00} \right] \delta h_{00} \\
			&+ \int d^{4}x \left[ - \frac{3}{4}  \omega \nabla^{2}  \partial_{i} \partial_{j} R + \frac{5}{4} \omega \delta^{ij} \nabla^{2} \nabla^{2} R - \omega \nabla^{4} R_{ij}  \right] \delta h_{ij} 
		\end{split}
	\end{equation}
	Using Eq. \eqref{deltahmunu}, we immediately observe that the BPS contribution vanishes because $\delta h_{00}=0$. Therefore, after substituting the canonical transformations, the variation of the action reduces to
	
	\begin{equation}
		\begin{split}
			\delta S & = \frac{1}{2} \int d^{4}x \left(   - \partial_{\lambda} \partial^{\lambda} h^{\mu \nu} + \eta^{\mu \nu} \partial_{\lambda} \partial^{\lambda} h^{\rho}_{\rho} -  \partial^{\mu} \partial^{\nu} h^{\lambda}_{\lambda} - \eta^{\mu \nu}  \partial^{\lambda} \partial^{\rho} h_{\lambda \rho} \right. \\
			& \left. +  \partial^{\mu} \partial^{\lambda}  h^{\nu}_{\lambda} +\partial^{\nu} \partial^{\lambda} h^{\mu}_{\lambda} \right) \left( \frac{1}{2} \left( \delta^{0}_{\mu} \delta^{i}_{\nu} + \delta^{0}_{\nu} \delta^{i}_{\mu} \right) \delta \omega_{2i} - \frac{1}{2} \delta^{i}_{\mu} \delta^{j}_{\nu} \left(  \partial_{i} \delta \omega_{4j} + \partial_{j} \delta \omega_{4i} \right) \right)  \\
			& + \int d^{4}x \left[ - \frac{3}{4}  \omega \nabla^{2}  \partial_{i} \partial_{j} R + \frac{5}{4} \omega \delta^{ij} \nabla^{2} \nabla^{2} R - \omega \nabla^{4} R_{ij}  \right] \left(  \partial_{i} \delta \omega_{4j} + \partial_{j} \delta \omega_{4i} \right)
		\end{split}
	\end{equation}
	In fact, we observe that after integrating by parts and using the linearized Bianchi identity, the contribution of the Cotton term vanishes identically. Therefore, although the Cotton term modifies the theory's dynamics through higher-order spatial derivatives, it preserves the gauge symmetry generated by the involutive Hamiltonians. Consequently, the  variation of the full action reduces to that of the $\lambda R$ sector,
	
	\begin{equation}
		\begin{split}
			\delta S_{\lambda R} & =  \frac{1}{2}  \int d^{4}x   \left[ \left( \nabla^{2} h_{0i} + \partial_{0} \partial_{i} h - \partial_{0} \partial_{j} h_{ij} - \partial_{i} \partial_{k} h_{0k} \right)  \left(  \delta \omega_{2i} + \partial_{0} \delta \omega_{4j} \right)  \right]. \\
		\end{split}
	\end{equation}
	Therefore, the complete action will be invariant if $\delta S=0$, thus, we find  the following transformation
	\begin{equation}
		\delta\omega_{2i} = -\partial_{0}\delta\omega_{4i}.
	\end{equation}
	Therefore, by identifying
	$\delta\omega_{4\mu}\equiv-2\omega_{\mu}$ and choosing
	$\omega_{0}=0$, the gauge transformations can be written as
	
	\begin{equation}
		\delta h_{\mu\nu}
		=
		\partial_{\mu}\omega_{\nu}
		+
		\partial_{\nu}\omega_{\mu}.
	\end{equation}
	This transformation coincides with the one obtained previously for the
	linearized $\lambda R$ model \cite{diana1}.\\
	This result also demonstrates how the gauge structure arises naturally within the Hamilton-Jacobi formalism. In this context, the involutive Hamiltonians play the role of first-class constraints in the Dirac formulation, while the generalized brackets account for the sector associated with second-class constraints. Consequently, the Hamilton-Jacobi analysis provides both the reduced dynamics and the gauge transformations without requiring a separate classification of the complete set of constraints at the end of the procedure.

	\section{The symplectic analysis for  $\lambda \neq \frac{1}{3} $}
	
	In the following sections, we will apply the FJ framework \cite{fad1, fad2, fad3, fad4}. From (\ref{Lag1}) we identify the symplectic Lagrangian given by 
	
	\begin{eqnarray}
		\nonumber
		\mathcal{L}^{(0)} & =&  \pi^{ij} \dot{h}_{ij} - G_{i j k l} \pi^{k l} \pi^{i j}+2 h_{j 0} \partial_{i} \pi^{i j}-\frac{1}{2} h_{00} R-\frac{1}{2} h^{i j}\left(R_{i j}- \frac{1}{2} \delta_{i j} R\right)+\alpha \partial_{i} h_{00} \partial^{i} h^{00}  \\
		& -& \omega \partial_{i} R_{k}^{j} \partial^{i} R_{j}^{k}+\omega \partial^{i} R_{i}^{j} \partial_{k} R_{j}^{k}+\frac{3 \omega}{8} \partial_{i} R \partial^{i} R+\frac{\omega}{2} \partial_{i} R_{j}^{i} \partial^{j} R,  
	\end{eqnarray}
	it has the form  $\mathcal{L}^{(0)}  = a_\alpha \dot{\xi}^\alpha- V^{(0)}$, for our case  $\mathcal{L}^{(0)}  =  \pi^{ij}\dot{h}_{ij}- V^{(0)}$, where the symplectic potential is given by 
	
	\begin{eqnarray}
		\nonumber 
		V^{(0)}&=& G_{i j k l} \pi^{k l} \pi^{i j}-2 h_{j 0} \partial_{i} \pi^{i j}+\frac{1}{2} h_{00} R+\frac{1}{2} h^{i j}\left(R_{i j}+ \frac{1}{2} \delta_{i j} R\right)-\alpha \partial_{i} h_{00} \partial^{i} h^{00}  \\
		& +& \omega \partial_{i} R_{k}^{j} \partial^{i} R_{j}^{k}-\omega \partial^{i} R_{i}^{j} \partial_{k} R_{j}^{k}-\frac{3 \omega}{8} \partial_{i} R \partial^{i} R-\frac{\omega}{2} \partial_{i} R_{j}^{i} \partial^{j} R,
	\end{eqnarray}
	thus the symplectic equations of motion are given by 
	
	\begin{equation}
		\label{eqmov}
		f_{\alpha \beta}^{(0)} \dot{\xi}^\beta= \frac{\partial V^{(0)}}{\partial \xi^ \alpha}, 
	\end{equation}
	where $f_{\alpha \beta}^{(0)}$ is called the symplectic matrix and it is given by 
	$$f^{(0)}_{\alpha\beta}(x,y) = \frac{\delta a_\beta(y)}{\delta \xi^\alpha(x)} - \frac{\delta a_\alpha(x)}{\delta \xi^\beta(y)}.$$
	From the symplectic Lagrangian, we identify the following  symplectic variables  given by $\xi^{\alpha} =(h_{ij}, h_{0i}, h_{00}, \pi^{ij} )$ and the one-form $a_{\alpha}=\left(\pi^{ij} ,0,0, 0\right) $, thus, the symplectic matrix is given by 
	
	\begin{equation}
		f_{\alpha\beta}(x,y) =  \begin{pmatrix} 0 & 0 & 0 & -I_{ij,kl} \\ 0 & 0 & 0 & 0 \\ 0 & 0 & 0 & 0 \\ I_{ij,kl} & 0 & 0 & 0 \end{pmatrix} \delta^3(x-y),
	\end{equation}
	where $I_{ij,kl} = \frac{1}{2}(\delta_i^k \delta_j^l + \delta_j^k \delta_i^l)$.  The null vectors of the symplectic matrix are given by 
	
	\begin{eqnarray}
		\mathcal{V}^{(1)}_k &=& (0, \delta_{k}^i, 0, 0), \\
		\mathcal{V}^{(2)} &=& (0, 0, 1, 0), 
	\end{eqnarray}
	from these null vectors, we obtain   by means of  \cite{fad1, fad2, fad3, fad4}
	
	\begin{equation}
		\label{cons}
		\Omega_k= \int dx^3 \mathcal{V}_k^{\alpha}\frac{\partial}{\partial \xi^{\alpha}(x)}\int V^{(0)}(\xi(y))dy^3, 
	\end{equation}
	the following constraints 
	
	\begin{eqnarray}
		\Omega^{(1)}_k &= & \partial_i\pi^{ik} = 0, \\
		\Omega^{(2)} &=&  \frac{1}{2}R + 2\alpha \nabla^2 h_{00} = 0.
	\end{eqnarray}
	Furthermore, we demand consistency conditions on these constraints,  similar to  Dirac's approach; this is 
	
	\begin{equation}
		\label{consis}
		\dot{\Omega}_k=\frac{\partial \Omega}{\partial \xi^\alpha}\dot{\xi}^{\alpha}=0,
	\end{equation}
	from the combination between (\ref{eqmov}) and (\ref{consis}), we obtain the following generalized system 
	
	\begin{eqnarray}
		f^{(1)}_{\alpha \beta} \dot{\xi}^{\beta}= Z_k(\xi), 
	\end{eqnarray}
	where 
	
	\begin{equation}
		\label{simp}
		f^{(1)}_{\alpha\beta} = \begin{pmatrix}  f^{(0)}_{\alpha\beta}& \\
			\frac{\partial \Omega_k}{\partial \xi^\alpha}
		\end{pmatrix},
	\end{equation}
	and 
	
	\begin{equation}
		Z_k(\xi)=  \begin{pmatrix}  \frac{\partial V^ {(0)}}{\partial \xi^\alpha}& \\
			0
		\end{pmatrix}.
	\end{equation}
	In this manner, we construct the symplectic matrix 
	
	\begin{equation}
		\label{sim1}
		f_{\alpha\beta}^{(1)}(x,y) =  \begin{pmatrix} 0 & 0 & 0 & -I_{ij,kl} \\ 0 & 0 & 0 & 0 \\ 0 & 0 & 0 & 0 \\ I_{ij,kl} & 0 & 0 & 0 \\
			0& 0 &0&I_{im,kl}\partial^m \\
			\partial^i \partial^j - \delta^{ij} \nabla^2& 0 &2\alpha \nabla^2 &0
		\end{pmatrix} \delta^3(x-y), 
	\end{equation}
	and 
	
	\begin{equation}
		Z_k(\xi) =  \begin{pmatrix}  \frac{\delta V^{(0)}}{\delta \xi^\alpha} \\  0  \end{pmatrix} = \begin{pmatrix}  \frac{\delta V^{(0)}}{\delta h_{ij}} \\  -2 \partial_j \pi^{ji} \\  \frac{1}{2} R + 2\alpha \nabla^2 h_{00} \\  2G_{ijkl} \pi^{kl} + \partial_i h_{0j} + \partial_j h_{0i} \\  0  \end{pmatrix},
	\end{equation}
	where $\frac{\delta V^{(0)}}{\delta h_{ij}} = \frac{1}{2} (\partial^i \partial^j - \delta^{ij}\nabla^2) h_{00} + R^{ij} - \frac{1}{2} \delta^{ij} R + \omega \nabla^4 R^{ij} + \frac{3}{4} \omega \nabla^2 \partial^i \partial^j R - \frac{5}{4} \omega \delta^{ij} \nabla^4 R$. The  matrix (\ref{sim1}) has the following null vectors 
	\begin{eqnarray}
		\mathcal{V}^{(1)}_k &=& (0, \delta_k^i, 0, 0, 0, 0), \\
		\mathcal{V}^{(2)} &=& (0, 0, 1, 0, 0, 0), \\
		\mathcal{V}^{(3)}_k &=& \left( -\frac{1}{2}(\delta_i^k \partial_j + \delta_j^k \partial_i), 0, 0, 0, \delta_m^k, 0 \right).
	\end{eqnarray}
	Thus, from the contraction of these null vectors with $Z_k(\xi)$, we obtain that  $\mathcal{V}_1^k Z_k=0$, therefore, no more constraints emerge. Now we add the constraints to the symplectic Lagrangian taking  the form
	\begin{eqnarray}
		\nonumber
		\mathcal{L}^{(2)} & =&  \pi^{ij} \dot{h}_{ij} +\dot{\rho}_j  \partial_{i} \pi^{i j}- \dot{\beta} \Big(  \frac{1}{2}R + 2\alpha \nabla^2 h_{00} \Big) - V^{(1)},
	\end{eqnarray}
	where  $V^{(1)}= V^{(0)}|_{\Omega_k}$.\\
	Now the symplectic variables are given by 
	$\xi^{\alpha} =(h_{ij}, \rho_i, h_{00}, \beta,  \pi^{ij} )$ and $a_{\alpha}=\left(\pi^{ij} ,  \partial_{j} \pi^{i j} ,0, \frac{1}{2}R + 2\alpha \nabla^2 h_{00}, 0 \right) $, and the symplectic matrix constructed with these symplectic variables  will be given by 
	\begin{equation}
		f^{(2)}_{\alpha\beta}(x,y) =  \begin{pmatrix} 0 & 0 & 0 & \frac{1}{2}(\partial_i \partial_j - \delta_{ij}\nabla^2) & -I_{ij,kl} \\ 0 & 0 & 0 & 0 & -I_{im,kl}\partial^m \\ 0 & 0 & 0 & 2\alpha \nabla^2 & 0 \\ -\frac{1}{2}(\partial_k \partial_l - \delta_{kl}\nabla^2) & 0 & -2\alpha \nabla^2 & 0 & 0 \\ I_{ij,kl} & I_{ij,km}\partial^m & 0 & 0 & 0 \end{pmatrix} \delta^3(x-y).
	\end{equation}
	We can observe that this matrix is still singular and has the following null vector $\mathcal{V}_k^{(3)} = \left( -I_{ij,kl}\partial^l, \delta_m^k, 0, 0, 0 \right)$, this null vector is the generator of gauge transformations; however, we have shown that there are no more constraints.  In this manner, for obtaining a symplectic tensor, we will fix the gauge by taking $h_{0i}=0$,  and  the symplectic Lagrangian now takes the form
	\begin{eqnarray}
		\nonumber
		\mathcal{L}^{(2)} =  \pi^{ij} \dot{h}_{ij} + \dot{\rho}_j \big( \partial_{i} \pi^{i j} +\sigma^j \big) - \dot{\beta} \Big(  \frac{1}{2}R + 2\alpha \nabla^2 h_{00} \Big)  -V^{1},
	\end{eqnarray}
	where we added $\sigma^i$ as Lagrange multiplier enforcing the gauge fixing. The new symplectic variables are identified as $\xi^{\alpha} =(h_{ij}, \rho_i, h_{00}, \beta,  \pi^{ij}, \sigma^i)$ and the one-form $a_{\alpha}=\left(\pi^{ij} ,  \partial_{j} \pi^{i j}+ \sigma^i,0, \frac{1}{2}R + 2\alpha \nabla^2 h_{00}, 0, 0 \right) $. Hence, with these symplectic variables we construct the  symplectic matrix 
	
	\begin{equation}
	\label{sym11}
		f^{(2)}_{\alpha\beta}(x,y) =  \begin{pmatrix} 0 & 0 & 0 & \frac{1}{2}(\partial_i\partial_j - \delta_{ij}\nabla^2) & -I_{ij,kl} & 0 \\ 0 & 0 & 0 & 0 & -I_{im,kl}\partial^m & -\delta_{ik} \\ 0 & 0 & 0 & 2\alpha\nabla^2 & 0 & 0 \\ -\frac{1}{2}(\partial_k\partial_l - \delta_{kl}\nabla^2) & 0 & -2\alpha\nabla^2 & 0 & 0 & 0 \\ I_{ij,kl} & I_{ij,km}\partial^m & 0 & 0 & 0 & 0 \\ 0 & \delta_{ik} & 0 & 0 & 0 & 0 \end{pmatrix} \delta^3(x-y),
	\end{equation}
	this matrix is not singular, and its  inverse is given by 
	
	\begin{equation}
	\label{sym1}
		(f^{(2)})^{\alpha\beta}(x,y) =  \begin{pmatrix} 0 & 0 & 0 & 0 & I_{ij,kl} & -I_{ij,km}\partial^m \\ \\ 0 & 0 & 0 & 0 & 0 & \delta_{ik} \\ \\ 0 & 0 & 0 & -\frac{1}{2\alpha\nabla^2} & -\frac{P_{kl}}{4\alpha\nabla^2} & 0 \\ \\ 0 & 0 & \frac{1}{2\alpha\nabla^2} & 0 & 0 & 0 \\ \\ -I_{ij,kl} & 0 & \frac{P_{ij}}{4\alpha\nabla^2} & 0 & 0 & 0 \\ \\ I_{im,kl}\partial^m & -\delta_{ik} & 0 & 0 & 0 & 0 \end{pmatrix} \delta^3(x-y), 
	\end{equation}
	from this matrix we can identify the following FJ brackets by means of   $\{ \xi^\alpha(x), \xi^\beta(y) \}_{FJ} = (f^{(2)})^{ \alpha\beta}(x,y)$, thus 
	
	\begin{align*}
		& \left\{h_{00}(x), \pi^{i j}(y)\right\}_{FJ}=-\frac{1}{4 \alpha \nabla^{2}} P^{i j}  \delta^{3}(x-y), \nonumber  \\
		& \left\{h_{i j}(x), \pi^{k l}(y)\right\}_{FJ}=\frac{1}{2}\left(\delta_{i}^{k} \delta_{j}^{l}+\delta_{j}^{k} \delta_{i}^{l}\right) \delta^{3}(x-y), 
	\end{align*}
	that corresponds to those found in (28) and (29) by means of HJ approach.\\
	On the other hand, we finish calculating  the functional measure for the path integral. In fact, in the FJ framework, we can compute the functional integration measure, which is determined by the square root of the determinant of the symplectic tensor \cite{Toms}.  Hence, we can calculate that determinant of (\ref{sym11}) by using (\ref{sym1}). For this aim, we observe that (\ref{sym1})  can be written as 
	
	\begin{equation}	
		(f^{(2)})^{\alpha\beta} = \begin{pmatrix} 0 & M^{-1} \\ -(M^{-1})^T & 0 \end{pmatrix}, 
	\end{equation}
	where we identify as  $M^{-1} = \begin{pmatrix} 0 & I_{ij,kl} & -I_{ij,km}\partial^m \\ 0 & 0 & \delta_{ik} \\ -\frac{1}{2\alpha\nabla^2} & -\frac{P_{kl}}{4\alpha\nabla^2} & 0 \end{pmatrix}$. In this manner,  after long computations, the determinant of $M^{-1}$ is given by  
	
	\begin{equation}
	\det(M^{-1}) = \left( -\frac{1}{2\alpha\nabla^2} \right) \det \begin{pmatrix} I_{ij,kl} & -I_{ij,km}\partial^m \\ 0 & \delta_{ik} \end{pmatrix}=  -\frac{1}{2\alpha\nabla^2}, 
	\end{equation}
	hence,  we find that $\det((f^{(2)})^{\alpha\beta}) = \frac{1}{4\alpha^2\nabla^4}$.  With this result at hand, we can identify the  functional measure \(d\mu\) for the partition function, it is given by 
	
	\begin{equation}
	d\mu = \prod_{\alpha} \mathcal{D}\xi^\alpha \sqrt{\det(f^{(2)}_{\alpha\beta})}.
	\end{equation}		
	Therefore, the measure for the theory under study will take the form $d\mu = \prod_{\alpha} \mathcal{D}\xi^\alpha (2\alpha\nabla^2) $, where we observe that it is $ \omega$-independent. Thus, the partition function can be found by 
	\begin{equation}
	Z = \int d\mu \exp\left( \frac{i}{\hbar} S_{\text{eff}} \right) = \int \left( \prod_{\alpha} \mathcal{D}\xi^\alpha \right) (2\alpha\nabla^2) \exp\left( i \int d^4x \, \mathcal{L}^{(2)} \right).
	\end{equation}
 Then, the path integral depends on the operator  $\nabla^2$ coupled by the   BPS $\alpha$ constant. This is a big difference between the case $\lambda\neq\frac{1}{3}$ and the critical one, because in the case of the critical stage, the measure will depend strongly on $ \omega$. We will see this in the following section. 

	\section{The symplectic analysis for $\lambda= \frac{1}{3} $}
	Now, from (\ref{Lag1}) we identify   the symplectic Lagrangian for the case $\lambda=\frac{1}{3}$ given by 
	
	\begin{eqnarray}
		\nonumber
		\mathcal{L}^{(0)} & = & \pi^{ij}\dot{h}_{ij} -\pi^{i j}  \pi_{i j}+2 h_{j 0} \partial_{i} \pi^{i j}-\frac{1}{2} h_{00} R-\frac{1}{2} h_{i j}\left(R^{i j}-\frac{1}{2} \delta^{i j} R\right)+\alpha \partial_{i} h_{00} \partial^{i} h^{00} \\
		& -&\omega \partial_{i} R_{k}^{j} \partial^{i} R_{j}^{k}+\omega \partial^{i} R_{i}^{j} \partial_{k} R_{j}^{k}+\frac{3 \omega}{8} \partial_{i} R \partial^{i} R+\frac{\omega}{2} \partial_{i} R_{j}^{i} \partial^{j}R + \dot{\mu} \pi,
	\end{eqnarray}
	it has the form  $\mathcal{L}^{(0)}  = a_\alpha \dot{\xi}^\alpha- V^{0}$, for our case  $\mathcal{L}^{(0)}  =  \pi^{ij}\dot{h}_{ij} + \dot{\mu} \pi- V^{(0)}$, where the symplectic potential is given by 
	
	\begin{eqnarray}
		\nonumber
		V^{(0)}&=& \pi^{i j}  \pi_{i j}-2 h_{j 0} \partial_{i} \pi^{i j}-\frac{1}{2} h_{00} R+\frac{1}{2} h_{i j}\left(R^{i j}+\frac{1}{2} \delta^{i j} R\right)-\alpha \partial_{i} h_{00} \partial^{i} h^{00} \\
		& +&\omega \partial_{i} R_{k}^{j} \partial^{i} R_{j}^{k}-\omega \partial^{i} R_{i}^{j} \partial_{k} R_{j}^{k}-\frac{3 \omega}{8} \partial_{i} R \partial^{i} R-\frac{\omega}{2} \partial_{i} R_{j}^{i} \partial^{j}R,
	\end{eqnarray}
	from the symplectic Lagrangian, we identify the symplectic variables  $\xi^{\alpha} =(h_{ij}, h_{0i}, h_{00},  \pi^{ij}, \mu)$ and the one-form $a_{\alpha}=\Big(\pi^{ij}, 0, 0, 0, \pi \Big) $, thus the symplectic matrix is given by 
	
	\begin{equation}
		f_{\alpha\beta}^{(0)}(x,y) =  \begin{pmatrix} 0 & 0 & 0 & -I_{ij,kl} & 0 \\ 0 & 0 & 0 & 0 & 0 \\ 0 & 0 & 0 & 0 & 0 \\ I_{ij,kl} & 0 & 0 & 0 & \delta_{ij} \\ 0 & 0 & 0 & -\delta_{kl} & 0 \end{pmatrix} \delta^3(x-y),
	\end{equation}
	this symplectic matrix has the following null vectors 
	
	\begin{eqnarray}
		\mathcal{V}_k^{(1)} &=& (0, \delta_k^i, 0, 0, 0), \nonumber  \\
		\mathcal{V}^{(2)} &=& (0, 0, 1, 0, 0),\nonumber  \\
		\mathcal{V}^{(3)} &= &(-\delta^{ij}, 0, 0, 0, 1),
	\end{eqnarray}
	from these null vectors and by using (\ref{cons}) we identify the following constraints
	
	\begin{eqnarray}
		\label{cons2} 
		\Omega_k^{(1)} &=& \partial_i\pi^{ik} = 0, \nonumber  \\
		\Omega^{(2)} &=&\frac{1}{2}R+ 2\alpha\nabla^2 h_{00} = 0, \nonumber \\
		\Omega^{(3)} &=&  \nabla^2 h_{00} + \frac{1}{2}R + 2\omega \nabla^4 R= 0.
	\end{eqnarray}
	In this manner, we will see if there are more constraints, by calculating  the $f^{(1)}_{\alpha\beta}(x,y)$ matrix given in (\ref{simp}), it is given by 
	
	\begin{equation}
		f^{(1)}_{\alpha\beta}(x,y) =  \begin{pmatrix} 0 & 0 & 0 & -I_{ij,kl} & 0 \\ 0 & 0 & 0 & 0 & 0 \\ 0 & 0 & 0 & 0 & 0 \\ I_{ij,kl} & 0 & 0 & 0 & \delta_{ij} \\ 0 & 0 & 0 & -\delta_{kl} & 0 \\ 
			0 & 0 & 0 & I_{ij,kl} \partial^l & 0 \\
			\frac{1}{2} (\partial^i \partial^j - \delta^{ij}\nabla^2) &0 & 2\alpha \nabla^2 & 0 & 0 \\
			\frac{\delta \Omega^{(3)}(x)}{\delta h_{ij}(y)} &0 &  \nabla^2 & 0 & 0 
		\end{pmatrix} \delta^3(x-y), 
	\end{equation}
	where 
	
	\begin{equation}
		\frac{\delta \Omega^{(3)}(x)}{\delta h_{ij}(y)} = \left[ \frac{1}{2}\partial_i \partial_j - \frac{1}{2}\delta_{ij}\nabla^2 + 2\omega \nabla^4 \partial_i \partial_j - 2\omega \delta_{ij}\nabla^6 \right] \delta^{(3)}(x-y),
	\end{equation}
	this matrix has the following null vectors
	
	\begin{eqnarray}
		\mathcal{V}_k^{(1)} &=& (0, \delta_k^i, 0, 0, 0, 0, 0, 0), \\
		\mathcal{V}^{(2)} &=& \left( -I_{ij,km}\partial^m \xi^k, 0, 0, 0, 0, \xi^m, 0, 0 \right), 
	\end{eqnarray}
	from these null vectors we don not find more constraints. In this manner, we will add the constraints (\ref{cons2}) to the symplectic Lagrangian
	
	\begin{eqnarray}
		\nonumber
		\mathcal{L}^{(1)} & =&  \pi^{ij} \dot{h}_{ij} +\dot{\mu}\pi  + \dot{\rho}_j \big( \partial_{i} \pi^{i j} \big)  - \dot{\beta} \Big(  \frac{1}{2}R + 2\alpha \nabla^2 h_{00}  \Big) \nonumber \\ 
		&+&\dot{\nu}\Big(  \nabla^2 h_{00} + \frac{1}{2}R + 2\omega \nabla^4 R \Big)- V^{(1)},  
	\end{eqnarray}
	where  $V^{(1)}= V^{(0)}|_{\Omega_k}$,  and the new symplectic variables are identified as  $\xi^{\alpha} =(h_{ij}, \rho_i, h_{00}, \beta,  \pi^{ij},   \mu, \nu)$ and the one-form $a_{\alpha}=\Big(\pi^{ij} ,  \partial_{j} \pi^{i j}, 0, \frac{1}{2}R + 2\alpha \nabla^2 h_{00}, 0, \pi,  \ \nabla^2 h_{00} + \frac{1}{2}R + 2\omega \nabla^4 R \Big) $. By using the symplectic variables, we construct the following symplectic matrix  
	
	\begin{equation}
		f_{\alpha\beta}(x,y) =  \begin{pmatrix} 0 & 0 & 0 & \frac{1}{2}(\partial_i\partial_j - \delta_{ij}\nabla^2) & -I_{ij,kl} & 0 & \frac{\delta {\Omega}^{(3)}}{\delta h_{kl}} \\ 0 & 0 & 0 & 0 & -I_{im,kl}\partial^m & 0 & 0 \\ 0 & 0 & 0 & 2\alpha\nabla^2 & 0 & 0 & \nabla^2 \\ -\frac{1}{2}(\partial_k\partial_l - \delta_{kl}\nabla^2) & 0 & -2\alpha\nabla^2 & 0 & 0 & 0 & 0 \\ I_{ij,kl} & I_{ij,km}\partial^m & 0 & 0 & 0 & \delta_{ij} & 0 \\ 0 & 0 & 0 & 0 & -\delta_{kl} & 0 & 0 \\ -\frac{\delta {\Omega}^{(3)}}{\delta h_{ij}} & 0 & -\nabla^2 & 0 & 0 & 0 & 0 \end{pmatrix} \delta^{(3)}(x-y),
	\end{equation}
	where    $\frac{\delta \Omega^{(3)}(x)}{\delta h_{ij}(y)} = \left[ \frac{1}{2}\partial_i\partial_j - \frac{1}{2}\delta_{ij}\nabla^2 + 2\omega \nabla^4 \partial_i\partial_j - 2\omega \delta_{ij}\nabla^6 \right]$. This matrix still is singular, in fact, it has the following null vector $\mathcal{V}_k^{} = \left( -I_{ij,km}\partial^m, \delta_k^i, 0, 0, 0, 0, 0 \right)$ and the gauge  symmetry generated by this vector is given by $h_{ij}\rightarrow h_{ij}- \partial_i \lambda_j-\partial_j \lambda_i$. 
	On the other hand, we find no additional constraints; hence, to obtain a symplectic tensor, we fix the gauge by taking $h_{0i}=0$. In this manner, the symplectic Lagrangian takes the form
	
	\begin{eqnarray}
		\nonumber
		\mathcal{L}^{(2)} & =&  \pi^{ij} \dot{h}_{ij} +\dot{\mu}\pi  + \dot{\rho}_j \big( \partial_{i} \pi^{i j} + \sigma^j \big)  - \dot{\beta} \Big(  \frac{1}{2}R + 2\alpha \nabla^2 h_{00}  \Big) \nonumber \\ 
		&+&\dot{\nu}\Big(  \nabla^2 h_{00} + \frac{1}{2}R + 2\omega \nabla^4 R \Big)- V^{(1)},  
	\end{eqnarray}
	where we have added the Lagrange multiplier $\sigma^i$ for enforcing the gauge fixing.  From the symplectic Lagrangian, we identify the following symplectic variables  $\xi^{\alpha} =(h_{ij}, \rho_i, h_{00}, \beta,  \pi^{ij},   \mu, \nu, \sigma^i)$ and $a_{\alpha}=\Big(\pi^{ij} ,  \partial_{j} \pi^{i j} + \sigma^i, 0, \frac{1}{2}R + 2\alpha \nabla^2 h_{00}, 0, \pi,  \ \nabla^2 h_{00} + \frac{1}{2}R + 2\omega \nabla^4 R \Big) $, hence, the symplectic matrix is given by 
	
	\begin{equation}
		f^{(2)}_{\alpha\beta}(x,y) =  \begin{pmatrix} 0 & 0 & 0 & \frac{1}{2}(\partial_i\partial_j - \delta_{ij}\nabla^2) & -I_{ij,kl} & 0 & \frac{\delta {\Omega}^{(3)}}{\delta h_{kl}} & 0 \\ 0 & 0 & 0 & 0 & -I_{im,kl}\partial^m & 0 & 0 & -\delta_{ik} \\ 0 & 0 & 0 & 2\alpha\nabla^2 & 0 & 0 & \nabla^2 & 0 \\ -\frac{1}{2}(\partial_k\partial_l - \delta_{kl}\nabla^2) & 0 & -2\alpha\nabla^2 & 0 & 0 & 0 & 0 & 0 \\ I_{ij,kl} & I_{ij,km}\partial^m & 0 & 0 & 0 & \delta_{ij} & 0 & 0 \\ 0 & 0 & 0 & 0 & -\delta_{kl} & 0 & 0 & 0 \\ -\frac{\delta {\Omega}^{(3)}}{\delta h_{ij}} & 0 & -\nabla^2 & 0 & 0 & 0 & 0 & 0 \\ 0 & \delta_{ik} & 0 & 0 & 0 & 0 & 0 & 0 \end{pmatrix} \delta^3(x-y),
	\end{equation}
	and its inverse is given by 
	
	\begin{equation}
	\label{sym2}
		(f^{(2)})^{\alpha\beta}(x,y) =  \begin{pmatrix} 0 & 0 & 0 & \frac{\delta_{ij}}{\nabla^2\Gamma} & I_{ij,kl} + \frac{\delta_{ij}P_{kl}}{2\nabla^2} & 0 & -\frac{2\alpha\delta_{ij}}{\nabla^2\Gamma} & -\frac{1}{2}(\partial_i\delta_{jk} + \partial_j\delta_{ik}) \\ \\ 0 & 0 & 0 & 0 & 0 & 0 & 0 & \delta_{ik} \\ \\ 0 & 0 & 0 & \frac{1+4\omega\nabla^4}{\nabla^2\Gamma} & 0 & 0 & -\frac{1}{\nabla^2\Gamma} & 0 \\ \\ -\frac{\delta_{kl}}{\nabla^2\Gamma} & 0 & -\frac{1+4\omega\nabla^4}{\nabla^2\Gamma} & 0 & 0 & -\frac{1}{\nabla^2\Gamma} & 0 & 0 \\ \\ -\left( I_{ij,kl} + \frac{P_{ij}\delta_{kl}}{2\nabla^2} \right) & 0 & 0 & 0 & 0 & -\frac{P_{ij}}{2\nabla^2} & 0 & 0 \\ \\ 0 & 0 & 0 & \frac{1}{\nabla^2\Gamma} & \frac{P_{kl}}{2\nabla^2} & 0 & -\frac{2\alpha}{\nabla^2\Gamma} & 0 \\ \\ \frac{2\alpha\delta_{kl}}{\nabla^2\Gamma} & 0 & \frac{1}{\nabla^2\Gamma} & 0 & 0 & \frac{2\alpha}{\nabla^2\Gamma} & 0 & 0 \\ \\ \frac{1}{2}(\partial_k\delta_{li} + \partial_l\delta_{ki}) & -\delta_{ik} & 0 & 0 & 0 & 0 & 0 & 0 \end{pmatrix} \delta^3(x-y)
	\end{equation}
	where $\Gamma = 1 - 2\alpha(1 + 4\omega\nabla^4)$. From this symplectic tensor,  we identify the  FJ brackets by means of   $\{ \xi^\alpha(x), \xi^\beta(y) \}_{FJ} = (f^{(2)})^{ \alpha\beta}(x,y)$, thus 
	
	\begin{equation}
		\left\{h_{i j}(x), \pi^{m n}(y)\right\}_{FJ}  =\frac{1}{2}\left(\delta_{i}^{m} \delta_{j}^{n}+\delta_{j}^{m} \delta_{i}^{n}\right) \delta^{3}(x-y)+\frac{\delta_{i j}}{2 \nabla^{2}} P^{m n} \delta^{3}(x-y),
	\end{equation}
	this corresponds to  (61) obtained  by means of the HJ approach. \\
	On the other hand, we finish calculating  the functional measure for  the path integral. In previous sections, we showed that in the FJ framework we can compute the functional integration measure for the path integral, which is determined by the square root of the determinant of the symplectic tensor (\ref{sym2}). For this aim, just like previous section, we reordering the symplectic tensor in the following way  $f_{\alpha \beta} = \begin{pmatrix} 0 & M \\ -M^T & 0 \end{pmatrix}$, where 
	
	\begin{equation}
		M  = \begin{pmatrix} -I_{ij,kl} & \frac{1}{2}P_{ij} & \tilde{\Omega}'_{ij} & 0 \\ 0 & 2\alpha\nabla^2 & \nabla^2 & 0 \\ -\delta_{kl} & 0 & 0 & 0 \\ 0 & 0 & 0 & \delta_{ik} \end{pmatrix}, 
	\end{equation}
	here $\tilde{\Omega}'_{ij} = \frac{\delta\Omega^{(3)}}{\delta h_{ij}}$ and  $P_{ij} = \partial_i\partial_j - \delta_{ij}\nabla^2$. Hence, the determinant will be given by $\det(f) = \det(M)^2$. After long computations, we obtain that the measure will take the form
	
	\begin{equation}
		d\mu = \prod_{\alpha} \mathcal{D}\xi^\alpha \sqrt{\det(f_{\alpha\beta})} = \prod_{\alpha} \mathcal{D}\xi^\alpha\left\{ \nabla^4 \left[ 1 - 2\alpha - 8\alpha\omega\nabla^4 \right] \right\},
	\end{equation}
	where we observe a difference with respect to the case $\lambda \neq \frac{1}{3}$  because the terms of the Cotton square  contribute at the quantum level. Finally, the partition function will be given by
	\begin{equation}
	Z = \int d\mu \, \exp\left( \frac{i}{\hbar} S^{(2)} \right) = \int \left( \prod_{\alpha} \mathcal{D}\xi^\alpha \right) \left\{ \nabla^4 \left[ 1 - 2\alpha - 8\alpha\omega\nabla^4 \right] \right\} \exp\left( i \int d^4x \, \mathcal{L}^{(2)} \right),
	\end{equation}
	where we observe a strong coupling  by $\omega$.\\
	We finish this section with the following comments. We note that because the lower-derivative infrared terms ($R$ and the BPS term $\alpha \partial_i h_{00}\partial^i h^{00}$) explicitly break anisotropic Weyl invariance, the Lagrangian (11) for $\lambda \neq 1/3$ represents a specific truncation of the general non-projectable Hořava theory, whereas for $\lambda = 1/3$ it represents a truncation of the critical kinetic-conformal theory with soft breaking of conformal symmetry by lower-dimensional operators. We previously noted that we chose the Cotton-squared term as a representative point of the $z=3$ ultraviolet potential with a single coupling parameter $\omega$. We could take a different sixth-order term in the Lagrangian; however, at the linearized level around Minkowski space, any parity-even sixth-order spatial curvature invariant (such as $\nabla_i R_{jk}\nabla^i R^{jk}$ and $\nabla_i R \nabla^i R$) reduces up to total derivatives to a linear combination of $\partial_i R_{jk}\partial^i R^{jk}$ and $\partial_i R \partial^i R$. Consequently, replacing the Cotton-squared term with the most general sixth-order spatial curvature potential does not qualitatively alter the canonical structure: for $\lambda \neq 1/3$, the constraints, generalized brackets, and path-integral measure remain completely independent of the sixth-order curvature terms, while for $\lambda = 1/3$, the trace constraint $H_6$ ($\Omega^{(3)}$) and the path-integral measure retain the exact same functional form up to a simple rescaling of the coupling $\omega$.
	\section{Conclusions}
	In this work, a detailed Hamilton-Jacobi analysis of the extended perturbative non-projectable $\lambda R$ model, incorporating the lowest-order BPS contribution and the Cotton tensor squared term, was performed. We obtained the complete set of Hamiltonians and constructed the corresponding fundamental differential for the cases $\lambda \neq 1/3$ and $\lambda = 1/3$.\\
 For the case $\lambda \neq 1/3$, the resulting fundamental bracket structure retains an explicit dependence on the BPS coupling $\alpha$ after the introduction of the fundamental HJ bracket. The constraint analysis reveals three physical degrees of freedom: two tensor modes and an additional scalar mode. In contrast, at the critical value $\lambda = 1/3$, the degeneracy of the generalized DeWitt metric gives rise to an additional constraint and modifies the non-involutive sector of the theory. Although the matrix associated with the non-involutive Hamiltonians depends explicitly on both $\alpha$ and $\omega$, this dependence cancels out in the resulting non-vanishing fundamental brackets. Furthermore, the theory propagates only two physical degrees of freedom, demonstrating that the additional scalar mode present in the generic case is absent when $\lambda = 1/3$. \\
 Additionally, while the BPS and Cotton tensor squared contributions modify the characteristic equations—and, consequently, the system's dynamics—they do not alter its gauge transformations. Thus, the gauge structure previously obtained for the perturbative $\lambda R$ model is preserved in the presence of both extensions. In this way, the additional operators modify the theory's dynamical content without altering its underlying gauge symmetry.\\
 From the symplectic perspective, we identified all constraints, constructed the symplectic tensor, and determined the path-integral measure. For $\lambda\neq\frac{1}{3}$, the constraints are $\omega$-independent, except for the nonlinear constraint $\Omega^{2}$, which depends on $\alpha$. This dependence is reflected in the symplectic tensor and affects the path integral measure. As a result, the quantum measure depends on the spatial momenta through the Laplacian and the BPS constant. \\
 On the other hand, in the critical case, we observed that the constraints depend on $\alpha$ and $ \omega$. In this case, higher-curvature terms contribute through $\omega$, and this contributes to the quantum measure. We have commented that  the Hořava model improves ultraviolet (UV) behavior by adding higher-order spatial derivatives without introducing higher-order time derivatives that cause instabilities; for the critical point, we observe that the Cotton tensor completely dominates the phase space topology in the high-energy regime because this is non-trivial and it is of eighth order in the spatial derivatives. In this manner, we observed that in the case of  $\lambda\neq\frac{1}{3}$ there are three degrees of freedom, and at the critical point there are two. In fact, at the critical point one degree of freedom is eliminated, but we pay a price because the measure now shows strong coupling through the constant $\omega$. Thus, we have all the tools to analyze the quantum aspects of the theory; the measure is already known, and we can use the   method presented in \cite{FAd} for such an analysis. Furthermore, using our results, we have already studied the complete Ho\v{r}ava theory, and the quantization process is in development and will be reported in forthcoming works \cite{Dianyo}.  \\
 Moreover, we add  the following final  remarks. In Ref. [16], applying Dirac's algorithm to this sixth-order spatial derivative theory required tracking primary, secondary, and tertiary constraints and classifying them into first- and second-class sets, leaving parts of the dynamical and quantum analysis incomplete. In contrast, we observed that the HJ formalism treats all constraints uniformly as Hamiltonians. It uses Frobenius integrability conditions to isolate the non-involutive sector immediately. At the same time, the FJ  symplectic method avoids introducing unphysical momenta ($\pi^{00}, \pi^{0i}$) altogether. It treats all constraints on equal footing via the null vectors of the symplectic matrix without any first- or second-class classification.\\
Unlike the Dirac approach in [16], we saw that the HJ fundamental differential directly yields the complete set of characteristic evolution equations for the phase-space variables (Eqs. 34–39 and 64–69). It extracts the exact gauge symmetries simply by setting $dt = 0$ (Section V), without requiring Castellani's algorithm.
Most importantly, the Dirac analysis in [16] did not determine the quantum path-integral measure. By constructing the invertible gauge-fixed symplectic tensor $f^{(2)}_{\alpha\beta}$ in the FJ framework, we directly compute the functional measure $d\mu = \prod_\alpha \mathcal{D}\xi^\alpha \sqrt{\det(f^{(2)}_{\alpha\beta})}$ in a single step. This reveals a key physical distinction between the two regimes that was not previously reported: although $\alpha$ and $\omega$ cancel out in the fundamental brackets $\{h_{ij}, \pi^{mn}\}^*$ at the critical point $\lambda = 1/3$, the Cotton-squared term enters non-trivially into the quantum path-integral measure for $\lambda = 1/3$ as an eighth-order spatial differential operator, $\nabla^4(1 - 2\alpha - 8\alpha\omega\nabla^4)$ (Eq. 123), whereas the measure for $\lambda \neq 1/3$ is completely independent of $\omega$ (Eq. 105), as was commented. Adopting the exact model of Ref. [16] is therefore essential to establish the equivalence of the generalized brackets across all three formalisms while demonstrating the clear advantages of the HJ and FJ methods for deriving the characteristic dynamics and the quantum partition function.


\end{document}